\documentclass[11pt,a4paper]{article}
\pdfoutput=1 %for arxiv

\usepackage{fullpage} 
\usepackage[tbtags]{amsmath}
\usepackage{amssymb}
\usepackage{amsthm}
\usepackage{mathtools}
\allowdisplaybreaks
\usepackage[T1]{fontenc}
\usepackage{lmodern}
\usepackage{bm}
\usepackage[hyphens]{url}
\usepackage{hyperref}
\usepackage[svgnames]{xcolor}
\hypersetup{colorlinks={true},urlcolor={DarkBlue},linkcolor={DarkBlue},citecolor=[named]{DarkGreen}}
\usepackage{microtype}
\usepackage[capitalise,nameinlink,noabbrev]{cleveref}
\usepackage[backend=biber, style=alphabetic, maxbibnames=999, maxalphanames=999, backref=true]{biblatex}
\usepackage{algorithm}
\usepackage{algpseudocode}

\usepackage{paralist}
\usepackage{mdframed}
\mdfsetup{backgroundcolor=white,nobreak=true}

\newtheorem{theorem}{Theorem}
\newtheorem{lemma}{Lemma}

\theoremstyle{definition}

\def \R{\mathbb R}
\def \N{\mathbb N}
\newcommand{\sset}[1]{\left\{ #1\right\}}
\newcommand{\ssets}[1]{\{ #1\}}
\newcommand{\fwh}[1]{\; \left| \; #1 \right.}
\newcommand{\card}[1]{\left| #1 \right|}

\newcommand{\prob}[1]{\ensuremath{\mathrm{Prob}\left[#1\right]}}

\DeclareMathOperator*{\argmin}{argmin}
\newcommand{\union}{\cup}
\newcommand{\map}{\longrightarrow}

\newcommand{\vecc}{\bm}
\DeclareMathOperator*{\expectation}{\mathbb E}
\newcommand{\expect}[2][]{\expectation_{#1}\nolimits\left[#2\right]}
\newcommand{\expectsmall}[2][]{\expectation_{#1}\nolimits[#2]}
\newcommand*{\poly}{\operatorname{\mathsf{poly}}}
\newcommand{\stochdom}{\mathrel{\leq_{\mathsf{st}}}}
\newcommand{\complexclass}[1]{\ensuremath{\mathsf{#1}}}

\title{A Smoothed FPTAS for Equilibria in Congestion Games}
\author{Yiannis Giannakopoulos\thanks{School of Computing Science, University of
Glasgow. Email: \href{mailto:yiannis.giannakopoulos@glasgow.ac.uk}{\nolinkurl
{yiannis.giannakopoulos@glasgow.ac.uk} }}}
\date{April 28, 2024}

\begin{document}
\maketitle

\begin{abstract}
We present a fully polynomial-time approximation scheme (FPTAS) for computing
equilibria in congestion games, under \emph{smoothed} running-time analysis. 
More precisely, we prove that if the resource costs of a congestion game are
randomly perturbed by independent noises, whose density is at most $\phi$, then
\emph{any} sequence of $(1+\varepsilon)$-improving dynamics will reach a
$(1+\varepsilon)$-approximate pure Nash equilibrium (PNE) after an expected
number of steps which is strongly polynomial in $\frac{1}{\varepsilon}$, $\phi$,
and the size of the game's description.
Our results establish a sharp contrast to the traditional worst-case analysis
setting, where it is known that better-response dynamics take exponentially long
to converge to $\alpha$-approximate PNE, for \emph{any} constant factor
$\alpha\geq 1$. As a matter of fact, computing $\alpha$-approximate PNE in
congestion games is \complexclass{PLS}-hard.

We demonstrate how our analysis can be applied to various different models of
congestion games including general, step-function, and polynomial cost, as well
as fair cost-sharing games (where the resource costs are decreasing). 
It is important to note that our bounds do not depend explicitly on the
cardinality of the players' strategy sets, and thus the smoothed FPTAS is
readily applicable to network congestion games as well.
\end{abstract}

\section{Introduction}

The systematic study of congestion games has its origins in the seminal work
of~\textcite{Rosenthal1973a}. Rosenthal, via a remarkably elegant construction,
proved that (unweighted) congestion games are potential
games~\parencite{Monderer1996a}, establishing in that way that they always have
pure Nash equilibria (PNE).
Since then, congestion games have been extensively studied in (algorithmic) game
theory and combinatorial optimization, since they provide a powerful abstraction
for modelling incentives in problems where different agents compete over a
 common collection of resources. 

From a computational perspective, the problem of computing a PNE of a congestion
game is a ``canonical'' local optimization problem, being a prominent member of
the complexity class \complexclass{PLS} introduced by~\textcite{Johnson:1988aa}.
As a matter of fact, as was first shown by~\textcite{FPT04}, the problem is
\complexclass{PLS}-complete. This hardness is two-fold. First, it implies that,
unless $\complexclass{P}=\complexclass{PLS}$, there does not exist an efficient
algorithm to compute equilibria in congestion games. Secondly, it proves that
better-response dynamics, which is simply the implementation of standard local
search in congestion games, can take exponentially long to converge to a PNE. It
is important to emphasize that the latter result is \emph{unconditional}, that
is, it does not depend on any complexity-theoretic assumptions.
\Textcite{Ackermann2008} showed that the \complexclass{PLS}-completeness is also
valid for network congestion games, that are defined succinctly over a graph
structure, and even for combinatorially very simple instances, with linear
resource cost functions.

Given the aforementioned hardness results, the natural direction to investigate
is the complexity of computing \emph{approximate} PNE. Unfortunately, it turns
out that the problem does not become easier: for \emph{any} given constant
$\alpha$, \textcite{Skopalik2008} showed that computing an $\alpha$-PNE is
\complexclass{PLS}-complete, and furthermore, proved the unconditional existence
of exponentially long better-response sequences, even for ``well-behaved''
resource costs.

Our goal in this paper is to demystify this dramatic complexity barrier, by
proving that the hard instances in congestion games are actually rather
``fragile''. To do to this formally, we deploy the framework of smoothed
analysis.

Smoothed analysis was first introduced in the groundbreaking work
of~\textcite{Spielman2004}, as a model for providing rigorous justification for
the empirical fact that the Simplex algorithm for linear programming, although
provably having exponential worst-case running-time, in practice  performs
exceptionally well. Their idea was very natural and remarkably effective: after an
input instance has been adversarially fixed, random perturbations are
introduced by ``nature'', independently, to all numerical parameters. Then, the
running-time of an algorithm is measured \emph{in expectation} with respect to
this randomness, termed \emph{smoothed running-time}. 

In the original model of~\parencite{Spielman2004}, the perturbations are
Gaussian around $0$, parameterized by their standard deviation $\sigma>0$; as
$\sigma\to 0$ this stochastic model converges to the original, fixed worst-case
instance. The seminal result of~\citeauthor{Spielman2004} says that the smoothed
running-time of Simplex (under the shadow-vertex pivot rule) is polynomial in
$\frac{1}{\sigma}$ (and the size of the input). One way to interpret this, is
that the ``bad'' instances for the performance of Simplex (see, e.g., the
Klee-Minty cube~\parencite{Klee1972}) are ``rare'' or ``isolated'', and
exponential precision is needed in their description in order to be effective.

Since~\parencite{Spielman2004}, smoothed analysis has been successfully applied
to a wide range of combinatorial problems, including, e.g., integer
programming~\parencite{Beier2006a,Roeglin2007a}, the $k$-means method for
clustering~\parencite{Arthur11}, multiobjective
optimization~\parencite{Brunsch15}, TSP~\parencite{Englert_2016,Englert14}, and
an impressive line of work on Local
Max-Cut~\parencite{ElsasserT11,Etscheid17,Angel:2017aa,Chen20,Bibak21}. As far
as game-theoretic problems are concerned, \textcite{Boodaghians20} studied the
smoothed complexity of finding PNE in network coordination games. 
Congestion games had not been studied from a smoothed analysis perspective
until very recently, when~\textcite{ggm2022_arxiv} showed that (exact) PNE can
be found in smoothed polynomial time for a (rather restrictive) class of games
that satisfy a certain ``constant-restraint'' assumption. 
For a more in-depth view of smoothed analysis we refer to, e.g.,
\parencite{Roughgarden2021,Spielman:2009aa,Beier2006a,ggm2022_arxiv}.
A more detailed presentation of the specific smoothness framework that we employ
in this paper is given in~\cref{sec:smoothed-congestion-games}.

We discuss further related work, in particular regarding various results about
the computability of approximate equilibria for the different models of
congestion games that we study in this paper, in the following, more 
technical sections.

\subsection{Our Results and Techniques}

In this paper we study the smoothed complexity of computing (approximate) pure
Nash equilibria (PNE) in (unweighted) congestion games. For our smoothed
analysis framework we follow the one recently proposed
in~\parencite{ggm2022_arxiv} for congestion games, where the cost of the
resources on the different possible loads are independently perturbed according
to an arbitrary probability distribution with density at most $\phi$. We
formalize our general congestion game model in~\cref{sec:model-notation}, where we
also define all necessary game-theoretic fundamentals.

In~\cref{sec:smoothed-FPTAS} we discuss (approximate) better-response dynamics
(BRD) and define our FPTAS (see~\cref{alg:approx-BRD}). Our main result is
stated in~\cref{th:main-FPTAS}: in general congestion games, a
$(1+\varepsilon)$-approximate PNE can be computed in \emph{smoothed} strongly
polynomial time in $\frac{1}{\varepsilon}$ (and the size of the game's
description). More precisely, $(1+\varepsilon)$-approximate BRD terminate after
at most $\tilde{O}(\varepsilon^{-1}\phi n^2 m^3)$ iterations (in expectation),
where $n$ is the number of players and $m$ the number of resources. The proof is
given in~\cref{sec:general-games-FPTAS} and the exact bound can be found
in~\eqref{eq:bound-smoothed-BRD-general-games-detailed}.

Furthermore, \cref{th:main-FPTAS} contains similar positive results for
additional, well-established special classes of congestion games. These differ
from general congestion games, in the way in which the resource cost functions
are defined and represented. Namely, we study: step-function costs with (a total
number of) $d$ break points; polynomial costs of constant degree $d$ (and
nonnegative coefficients); and fair cost-sharing games where a fixed cost is
equally split among the players who use it. The corresponding smoothed
complexity bounds on the number of iterations of $(1+\varepsilon)$-BRD are,
respectively: $\tilde{O}(\varepsilon^{-1}\phi n m d^2)$
(see~\cref{sec:step-function-FPTAS} for the proof),
$\tilde{O}(\varepsilon^{-1}\phi n^{d+1} m^3)$ (\cref{sec:polynomial-FPTAS}), and
$\tilde{O}(\varepsilon^{-1}\phi n m^3)$ (\cref{sec:cost-sharing-FPTAS}).

It is worth mentioning that the aforementioned bounds hold for any starting
configuration of the dynamics, and for any choice of the intermediate pivoting
rule for the player deviations. Furthermore, all our results are immediately
valid for \emph{network} congestion games (see~\cref{sec:congestion-model} for a
definition and~\cref{sec:smoothed-FPTAS} for a discussion) as well, since the
running time of our FPTAS does not depend on the number of strategies available
to the players (which, for network games, can be exponential in $n$ and $m$).

The technique for achieving the smoothed polynomial complexity bounds in this
paper can be distilled in two core steps. First, we establish that the number of
iterations of BRD can be upper bounded by an appropriate function of the ratio
between the maximum and minimum resource costs of our game; for general
congestion games, for example, this can be seen
in~\eqref{eq:bound-BRD-smoothed-ratio-values}. Similar inequalities hold for the
other special congestion game models that we study, and they all arise from the
algebraic relation between player costs and the value of Rosenthal's potential
(see~\cref{sec:congestion-model} for definitions). 

Secondly, we show that when this expression is paired with a simple
(exponential, based on exhaustive search) bound on the running time
(see~\eqref{eq:bound-smoothed-BRD-general-games-helper-1}), the expectation of
the resulting quantity grows polynomially. This probabilistic property is the
cornerstone for our derivation, and we present it in its
own~\cref{sec:prob-lemma}, before we dive into the rest of the technicalities in
our proofs. The presentation in~\cref{sec:prob-lemma} is essentially
self-contained, independent of congestion games, and
\cref{lemma:smooth-minimum-order-statistic-bound}  applies to general
$\phi$-smooth random variables. As a result, it may prove useful for future work
in smoothed analysis, whenever similar bounds involving the ratios of the
numerical parameters of the problem can be shown to hold.

\section{Model and Notation}
\label{sec:model-notation}

We will use $\N$, $\R$, and $\R_+$ to denote the set of nonnegative integer,
real, and nonnegative real numbers, respectively. For $n\in\N$ we denote
$[n]\coloneqq\ssets{1,2,\dots,n}$ and $[0..n]\coloneqq\ssets{0}\union[n]$.
For a random variable $X$ we use $F_X$ for its cumulative distribution function
(cdf) and $f_X$ for its probability density function (pdf). In this paper we
only deal with (absolutely) continuous, real-valued random variables. 
We will use $\stochdom$ for the usual (first-order) stochastic ordering; that
is, for two random variables $X,Y$: $X \stochdom Y$ if and only if $F_Y(t) \leq
F_X(t)$ for all $t\in\R$.

\subsection{Congestion Games}
\label{sec:congestion-model}

A \emph{congestion game} $\mathcal{G}=\left(N,R,\ssets{S_i}_{i\in
N},\ssets{c_r}_{r\in R}\right)$ is defined by 
\begin{inparaenum}[(1)]
    \item a finite set of \emph{players} $N=[n]$,
    \item a finite set of \emph{resources} $R=\ssets{r_1,r_2,\dots,r_m}$,
    \item for each player $i$, a \emph{strategy set} $ S_i\subseteq
    2^R\setminus\ssets{\emptyset}$; each element $s_i\in S_i$ is thus a nonempty
    set of resources, and is called a \emph{strategy} for player $i$, and
    \item for each resource $r\in R$, a \emph{cost function} $c_r:[n]\map\R_+$;
    $c_r(\ell)$ is interpreted as the cost (or congestion) of resource $r$ when
    $\ell$ players use it.
\end{inparaenum}

A \emph{network congestion game} is a congestion game where the strategy sets
$\ssets{S_i}_{i\in N}$ are not given explicitly, but induced via an underlying
graph structure. More precisely, we are given a directed graph $G=(V,E)$ and,
for each player $i\in N$ a pair of nodes $(o_i,d_i)\in V$. The resources of the
game are exactly the edges of the graph, i.e.\ $R=E$. Then, the strategies of
player $i$ are all simple $o_i\to d_i$ paths in $G$.

Notice how, in the definition above, we do not enforce any monotonicity
requirement on the resource cost functions, since our main result does not
depend on such an assumption and applies to general congestion games with
arbitrary cost functions (see case (\ref{item:main-FPTAS-general})
of~\cref{th:main-FPTAS} and the corresponding proof
in~\cref{sec:general-games-FPTAS}).
We discuss more specialized congestion game models, including step-function and
polynomial costs (which are nondecreasing) and cost-sharing games (where the
costs are decreasing) in their corresponding
\cref{sec:step-function-FPTAS,sec:polynomial-FPTAS,sec:cost-sharing-FPTAS}. In
such models, the resource costs $c_{r}(\ell)$ are not given explicitly, but
rather via a more succinct functional expression.

A \emph{strategy profile} of a congestion game $\mathcal{G}$ is a collection of
strategies, one for each player: $\vecc{s}=(s_1,s_2,\dots,s_n)\in
\vecc{S}\coloneqq S_1\times S_2\times\dots\times S_n$. For a strategy profile
$\vecc s$, we use $\ell_r(\vecc s)$ for the \emph{load} it induces on a resource
$r\in R$, that is, the number of players that use it: $\ell_r(\vecc s)\coloneqq
\card{\sset{i\in N\fwh{r\in s_i}}}$. This induces a cost to the players, equal
to the sum of the cost of the resources that they are using. That is, the
\emph{cost of player} $i\in N$ under a strategy profile $\vecc s\in \vecc{S}$ is:
$$
C_i(\vecc s) \coloneqq \sum_{r\in s_i} c_r(\ell_r(\vecc s)).
$$

For an $\alpha\geq 1$, we will say that a strategy profile is an
\emph{$\alpha$-approximate pure Nash equilibrium ($\alpha$-PNE)} if no player
can improve their cost more than a (multiplicative) factor of $\alpha$, by
unilaterally deviating to another strategy. Formally, $\vecc s\in\vecc{S}$ is an
$\alpha$-PNE if\footnote{Here we are using the standard game-theoretic notation
of $\vecc{s}_{-i}$ to denote the $(n-1)$-dimensional vector that remains from
the $n$-dimensional vector $\vecc{s}$ if we remove its $i$-th coordinate. In
that way, for any vector $\vecc s$ we can write $\vecc s=(s_i,\vecc{s}_{-i})$.}
\begin{equation}
    \label{eq:PNE-def}
C_i(\vecc s) \leq \alpha \cdot C_i(s_i',\vecc{s}_{-i}),
\qquad \text{for all}\;\; i\in N,\; s_i'\in S_i.
\end{equation}
For the special case of $\alpha=1$ this definition coincides with the stricter,
standard notion of a pure Nash equilibrium. To emphasize this, sometimes a
$1$-PNE is called an \emph{exact} PNE. Notice that any exact
PNE is also an $\alpha$-PNE, for any $\alpha\geq 1$.

The \emph{Rosenthal potential} of a congestion game $\mathcal G$ is the function
$\Phi:\vecc{S}\map\R_+$ given by
\begin{equation}
    \label{eq:Rosenthal-potential}
    \Phi(\vecc s) \coloneqq \sum_{r\in R}\sum_{j=1}^{\ell_r(\vecc s)} c_r(j).
\end{equation}
This is due to the work of \textcite{Rosenthal1973a} who first defined the
quantity in~\eqref{eq:Rosenthal-potential} and proved that, for all strategy
profiles $\vecc s$ of a congestion game we have
\begin{equation}
    \label{eq:potential-equals-cost-diff}
\Phi(\vecc s) - \Phi(s_i',\vecc{s}_{-i}) = C_i(\vecc s) - C_i(s_i',\vecc{s}_{-i}).
\end{equation}
An immediate consequence of~\eqref{eq:potential-equals-cost-diff}, also shown
by~\textcite{Rosenthal1973a}, is that a minimizer of Rosenthal's potential
$\vecc s^{*}\in\argmin_{\vecc s \in\vecc{S}} \Phi(\vecc s)$ is an exact PNE.
This establishes the existence of $\alpha$-PNE (for any $\alpha\geq 1$) in all
congestion games.

The goal of the present paper is to study computationally efficient methods for
computing such an $\alpha$-PNE, for a factor $\alpha$ as close to $1$ as
possible.

\subsubsection{Smoothed Congestion Games}
\label{sec:smoothed-congestion-games}

In this paper we study the complexity of our algorithms under \emph{smoothed
analysis} (see, e.g., \parencite[Part~4]{Roughgarden2021}). 
In a \emph{$\phi$-smooth congestion game}, we assume that the resource costs
$\ssets{c_r(\ell)}_{r\in R,\ell\in[n]}$ are independent
random variables; then, we measure running-times \emph{in expectation} with
respect to their realizations. In more detail, we assume that the resource costs
are continuous random variables, taking values in $[0,1]$, and that their
density functions are upper-bounded by a universal parameter $\phi\geq 1$. More
generally, we will call such a random variable $X$ with $f_X:[0,1]\map
[0,\phi]$, a \emph{$\phi$-smooth random variable}. 

Notice here that the normalization of the
resource costs within $[0,1]$ is without loss for our purposes: one can just
divide all costs by their maximum, and get a totally equivalent game that fully
maintains the equilibrium structure. Such a scaling is done to facilitate
the smoothed analysis modelling, and is standard in the field (see, e.g.,
\parencite{Englert_2016,Etscheid17,ggm2022_arxiv}).

Parameter $\phi$ allows smoothed analysis to interpolate between average-case
analysis ($\phi=1$) where all costs are uniformly i.i.d., and worst-case
analysis ($\phi=\infty$) where the $\phi$-smooth resource costs degenerate to
single values. The aim of smoothed analysis is to capture the complexity of an
algorithm, asymptotically as $\phi$ grows large. To see it from another
perspective, smoothed analysis can be seen as introducing small, independent,
random perturbations to the numerical values of a problem instance,
\emph{before} performing a traditional, worst-case running-time analysis. The
magnitude of this random noise can be ``controlled'' by a parameter
$\sigma=\frac{1}{\phi}\to 0$.
A detailed discussion of the fundamentals and subtleties of smoothed analysis is
beyond the scope of this paper; for such a treatment, the interested reader is
referred to, e.g., \parencite{Roughgarden2021,Spielman2004,Beier2006a}.

It is important to clarify that, in this section we introduced our smoothness
framework only for general congestion games. Due to its nature, smoothed
analysis depends heavily on the representation and the numerical parameters of
each problem instance; therefore, different special models of congestion games
require their own, tailored smoothness treatment. To assist readability, we have
decided to defer the discussion of smoothness for step-functions, polynomial,
and fair cost-sharing games to their
corresponding~\cref{sec:step-function-FPTAS,sec:polynomial-FPTAS,sec:cost-sharing-FPTAS}.
For most of these models, we adopt the smoothness frameworks for congestion
games that were first proposed recently by~\parencite{ggm2022_arxiv}; except
for cost-sharing games (see~\cref{sec:cost-sharing-FPTAS}) for which, to the
best of our knowledge, this is the first time that they've been studied from a
smoothed analysis perspective.

\section{The Smoothed FPTAS}
\label{sec:smoothed-FPTAS}

In this section we present our FPTAS (see~\cref{alg:approx-BRD}). It is a based
on a very simple, but fundamental idea. Fix an arbitrary game $\mathcal G$ and a
parameter $\alpha\geq 1$. If a strategy profile $\vecc s$ of $\mathcal G$ is
\emph{not} an $\alpha$-approximate PNE, then (by simply considering the negation
of~\eqref{eq:PNE-def}) there has to exist a player $i$ and a strategy $s_i'$ of
$i$ that improves their cost by a factor larger than $\alpha$; formally:
\begin{equation}
    \label{eq:alpha-improving-move}
    \alpha C_i(s_i',\vecc s_{-i}) < C_i(\vecc s)
\end{equation}
Such a deviation $\vecc{s} \to (s_i',\vecc{s}_{-i})$, that
satisfies~\eqref{eq:alpha-improving-move}, is called an \emph{$\alpha$-improving
move} for game $\mathcal G$. 

This gives rise to the following natural process for finding approximate
equilibria in games, called \emph{$\alpha$-better-response dynamics
($\alpha$-BRD)}: starting from an arbitrary strategy profile, repeatedly perform
$\alpha$-improving moves. When no such move exists any more, it must be that  an
$\alpha$-PNE has been reached. For a more formal description,
see~\cref{alg:approx-BRD}.
\begin{algorithm}[t]
    \caption{$\alpha$-Approximate Better-Response Dynamics ($\alpha$-BRD),
    $\alpha\geq 1$}\label{alg:approx-BRD}
    \begin{algorithmic}[1]
    \Require Congestion game $\mathcal G=\left(N,R,\ssets{S_i}_{i\in
    N},\ssets{c_r}_{r\in R}\right)$; strategy profile $\vecc s\in \vecc S$
    \Ensure An $\alpha$-PNE of $\mathcal G$ \While{$\vecc s$ is not an
    $\alpha$-PNE} \label{alg_line:BRD_check-equilibrium} \State Choose $i\in N$,
    $s_i'\in S_i$ such that: $\alpha C_i(s_i',\vecc s_{-i}) < C_i(\vecc s)$
    \label{alg_line:BRD_improvement-step} \State $\vecc s \gets (s_i',\vecc
    s_{-i})$ \label{alg_line:BRD_update-configuration} \EndWhile \State\Return
    $\vecc s$
    \end{algorithmic}
\end{algorithm}

Notice how in Line~\ref{alg_line:BRD_improvement-step} of \cref{alg:approx-BRD}
there might be multiple valid $\alpha$-improving moves $\vecc s \to
(s_i',\vecc{s}_i)$ to choose from. Our definition deliberately leaves this
underdetermined, as all the results presented in this paper hold for  \emph{any}
choice of the $\alpha$-improving moves. Furthermore, we have made the starting
profile to be part of the input, in order to emphasize the fact that this can be
adversarially selected; again, all our bounds are robust to the choice of an
initial configuration.

The main result of our paper is that, under smoothed running-time analysis,
approximate better-response dynamics converge fast to approximate equilibria:
\begin{mdframed}[backgroundcolor=gray!30]
\begin{theorem}
    \label{th:main-FPTAS}
    In $\phi$-smooth congestion games, $(1+\varepsilon)$-BRD always find a
    $(1+\varepsilon)$-PNE, after an expected number of iterations which is
    strongly polynomial in $\frac{1}{\varepsilon}$, $\phi$, and the description
    of the game. This holds for
    \begin{inparaenum}[(a)]
        \item\label{item:main-FPTAS-general} general,
        \item\label{item:main-FPTAS-step-functions} step-function,
        \item\label{item:main-FPTAS-polynomial} polynomial, and
        \item\label{item:main-FPTAS-cost-sharing} cost-sharing
    \end{inparaenum}
    congestion games, and even under a succinct network representation of all
    models
    (\ref{item:main-FPTAS-general})--(\ref{item:main-FPTAS-cost-sharing}). 
    
    More precisely, the expected number of iterations is at
    most $\left(1+\frac{1}{\varepsilon}\right)\poly(\phi,n,m)$ where $n$ is the
    number of players and $m$ the number of resources.
\end{theorem}
\end{mdframed}

Notice that \cref{th:main-FPTAS} gives a bound on the number of
\emph{iterations} of our dynamics (i.e., the while-loop of
Lines~\ref{alg_line:BRD_check-equilibrium}--\ref{alg_line:BRD_update-configuration}),
and not the \emph{total running time}. This is due to fact that, checking
whether a given strategy profile is an (approximate) equilibrium
(Line~\ref{alg_line:BRD_check-equilibrium} of~\cref{alg:approx-BRD}) and, if
not, returning an improving move (Line~\ref{alg_line:BRD_improvement-step}), can
both be done in polynomial time (in the description of the game), for all
congestion game models studied in this paper; therefore, the total running time
is indeed dominated by the number of improving-move steps.

To see that indeed that's the case, first consider the standard representation
of congestion games, where the strategy sets $\ssets{S_i}_{i\in N}$ of the
players are \emph{explicitly} given in the input, each $S_i$ being a list of, at
most $k$, subsets of elements of the ground set of resources $R$. Then, one can
actually compute \emph{all} $\alpha$-improving moves by simply exhaustively
going over all players $i\in N$, and all strategy deviations $s_i'\in S_i$,
checking whether $\frac{C_i(\vecc s)}{C_i(s_i',\vecc{s}_{-i})}>\alpha$; this can
be done in $O(nk)$ time.
On the other hand, for congestion game representations where the strategy sets
are \emph{implicitly} given, it might not be possible to efficiently perform
such an exhaustive search. Then, one needs to have access to the fundamental
game-theoretic primitive of a \emph{best-response}, i.e., for every player $i$
and any strategy profile $\vecc{s}_{-i}\in \vecc{S}_{-i}$, being able to
efficiently compute an element
\begin{equation}
    \label{eq:BR-set-def}
s_i'\in\argmin_{s_i\in S_i}\sset{C_i(s_i,\vecc{s}_{-i})}.
\end{equation}
Then, finding an $\alpha$-improving move can still be done in polynomial time,
since it boils down to going over the players $i\in N$, computing a
best-response $s_i'$, and checking whether $\frac{C_i(\vecc
s)}{C_i(s_i',\vecc{s}_{-i})}>\alpha$. If this check fails for all players, then
it must be that $\vecc s$ is an $\alpha$-PNE.
In particular, this applies to network congestion games (recall the definition
from~\cref{sec:congestion-model}) where each strategy set $S_i$ is succinctly
described as a set of paths between two fixed nodes, and therefore its
cardinality might be exponential. However, for such games, best-responses
in~\eqref{eq:BR-set-def} are simply shortest-path computations, and thus they
can indeed be performed efficiently.

\bigskip

The rest of our paper is devoted to proving~\cref{th:main-FPTAS} and describing
the smoothness frameworks for all the different congestion game models
(\ref{item:main-FPTAS-general}), (\ref{item:main-FPTAS-step-functions}),
(\ref{item:main-FPTAS-polynomial}), and (\ref{item:main-FPTAS-cost-sharing})
that are involved in the statement of~\cref{th:main-FPTAS} and which we study in
this paper.

\subsection{The Key Probabilistic Lemma}
\label{sec:prob-lemma}

A critical step in the proof of our main results (\cref{th:main-FPTAS}) is that
of identifying a key property about $\phi$-smooth random variables. In order to
highlight its importance, we decided to disentangle it from the rest of the
running-time analysis of our FPTAS, and present it beforehand here in its own
section, together with its proof
(see~\cref{lemma:smooth-minimum-order-statistic-bound}). Furthermore, since it
is completely independent of congestion games, it can be of particular interest
for future work on smoothed analysis. 

To provide intuition first, consider some combinatorial optimization problem
involving numerical inputs $w_1,w_2,\dots,w_m$, normalized in $(0,1]$. For
example, these could be the weights of a knapsack instance, or the coefficients
of an integer program. Then, $W=\max_i \frac{1}{w_i}=\frac{1}{\min_i w}_i$
effectively captures the ``magnitude'' of the numbers that are involved in our
computation. A running time which is polynomial on $W$ does not, in general,
imply efficient computation; for example, this is highlighted by many NP-hard
problems that are known to admit pseudopolynomial solutions (like knapsack, for
example). Nevertheless, what if, under the more optimistic lens of smoothed
analysis, one could show that the magnitude of $W$ is ``well-behaved'' with
``high probability''?

The following observation immediately shatters such hopes: consider a uniformly
distributed random variable $X$ over $[0,1]$, and observe that the expectation
$\expectsmall{\frac{1}{X}}=\int_{0}^1\frac{1}{x}\, \mathrm{d}x=\infty$ of its
reciprocal is actually unbounded. Nevertheless, it turns out that a small patch
is enough to do trick: truncating the random variable $\frac{1}{X}$ from above,
even at an  exponentially large threshold, results in a polynomially bounded
expectation. This is formalized in the following:

\begin{mdframed}
\begin{lemma}
    \label{lemma:smooth-minimum-order-statistic-bound} Let $X_1,X_2,\dots,X_\mu$
    be independent $\phi$-smooth random variables over $[0,1]$. For any reals
    $\alpha\geq 1$, $\beta\geq 0$ it holds that:
    \begin{equation}
        \label{eq:smooth-minimum-order-statistic-bound}
        \expect{\min\sset{\max_{i\in[\mu]} \frac{\alpha}{X_i},\mu^\beta}} 
        \leq \phi \alpha (\beta+1) \mu\ln{\mu} + 1
        = \tilde{O}(\phi \alpha\beta\mu).
    \end{equation}
\end{lemma}
\end{mdframed}
\begin{proof}
    To simplify notation, we first define function $g:[0,1]\map\R$ with
    $$g(t)\coloneqq \min\sset{\frac{\alpha}{t},\mu^\beta}.$$
    Notice that function $g$ is nonincreasing and that the expectation
    in~\eqref{eq:smooth-minimum-order-statistic-bound} can now be more simply
    expressed as $\expect{g(\min_i X_i)}$.

    Now let $Y_1, Y_2,\dots, Y_\mu$ be independent uniformly distributed random
    variables over $[0,\frac{1}{\phi}]$. Notice that they are $\phi$-smooth and
    their cdf is given by $F_{Y}(y)=\phi y$ for $0\leq y \leq \frac{1}{\phi}$.
    Let us also define the random variable $Z\coloneqq \frac{1}{\min_{i\in[\mu]}
    Y_i}$. Observe that $Z$ takes values in $[\phi,\infty)$ and its cdf and pdf
    are given by
    \begin{align*}
    F_Z(z) &=\prob{Z\leq z}
    =\prod_{i=1}^\mu\prob{Y_i\geq \frac{1}{z}}
    =\prod_{i=1}^\mu\left[1-F_{Y}\left(\frac{1}{z}\right)\right]
    =\left(1-\frac{\phi}{z}\right)^\mu,
    \end{align*}
    and
    \begin{align*}
        f_Z(z) &=\frac{\mathrm{d}\, F_Z(z)}{\mathrm{d}\, z}
            = \mu \left(1-\frac{\phi}{z}\right)^{\mu-1}\frac{\phi}{z^2},
        \end{align*}
    respectively, for $z\geq \phi$. 
    From these expressions, for all $z\geq \phi$ we furthermore get the bounds
    \begin{equation}
        \label{eq:smooth-minimum-order-statistic-bound-helper-pdf-cdf-bounds}
        F_Z(z) \geq 1-\frac{\mu\phi}{z} \qquad\text{and}\qquad f_Z(z) \leq \frac{\mu\phi}{z^2},
    \end{equation}
    where the first inequality can be derived by using Bernoulli's
    inequality.\footnote{Bernoulli's inequality (for a proof see, e.g.,
    \textcite[\S{0.2}]{Mitrinovic1964}) states that, for all positive integers
    $m$ and all reals $y\geq -1$ it holds that $(1+y)^m \geq 1+my $.
    Instantiating this with $y \gets -\frac{\phi}{z}\geq -1 $ and $m\gets \mu$
    we get the desired first inequality
    in~\eqref{eq:smooth-minimum-order-statistic-bound-helper-pdf-cdf-bounds}.}
    
    Next, we argue that $Y_i \stochdom X_i$, where $\stochdom$ denotes the usual
    (first-order) stochastic order. Indeed, since each $X_i$ is $\phi$-smooth,
    its cdf is upper-bounded by
    $$
    F_{X_i}(x) \leq \int_0^x \phi \, \mathrm{d}t =\phi x = F_Y(x),
    $$
    for all $0\leq x \leq \frac{1}{\phi}$; obviously, $F_{X_i}(x)\leq 1 = F_Y(x)$ for all $x\geq \frac{1}{\phi}$ as well. 
    Since order statistics preserve stochastic dominance (see, e.g.,
    \cite[p.~100]{Boland1998}), it must be that $\min_{i} Y_i \stochdom \min_{i}
    X_i$. Thus, by the monotonicity of function $g$ it must be that (see, e.g.,
    \cite[Proposition~9.1.2]{Ross1996})
    \begin{equation}
        \label{eq:smooth-minimum-order-statistic-bound-helper-1}
        \expect{g\left(\min\nolimits_i X_i\right)} \leq \expect{g\left(\min\nolimits_i Y_i\right)}= \expect{\min\ssets{\alpha Z ,\mu^\beta}}.
    \end{equation}
    
    Using this, and denoting $z^*\coloneqq \phi\mu^{\beta+1} \geq \phi$ for convenience, we can finally upper bound the expectation in~\eqref{eq:smooth-minimum-order-statistic-bound}
    by:
    \begin{align*}
        \expect{g\left(\min\nolimits_i X_i\right)} 
        &= \int_{\phi}^\infty \min\ssets{\alpha z,\mu^\beta} f_Z(z)\,\mathrm{d}z\\
        &\leq \int_{\phi}^{z^*} \alpha z f_Z(z)\,\mathrm{d}z
        + \int_{z^*}^\infty \mu^\beta f_Z(z)\,\mathrm{d}z\\
        &= \alpha\int_{\phi}^{z^*} z f_Z(z)\,\mathrm{d}z
        + \mu^\beta \left[1-F_Z(z^*)\right]\\
        &\leq \alpha\int_{\phi}^{z^*} z \frac{\mu\phi}{z^2}\,\mathrm{d}z
        + \mu^\beta \left[1-\left(1-\frac{\mu\phi}{z^*}\right)\right], &&\text{by using~\eqref{eq:smooth-minimum-order-statistic-bound-helper-pdf-cdf-bounds},}\\
        &= \alpha\mu\phi\ln\frac{z^*}{\phi} + \frac{\mu^{\beta+1}\phi}{z^*}\\
        &= \alpha\mu\phi(\beta+1)\ln\mu + 1, &&\text{since}\;\; z^*=\phi\mu^{\beta+1}.
    \end{align*}
    \end{proof}

\subsection{General Congestion Games}
\label{sec:general-games-FPTAS}

In this section we prove part~(\ref{item:main-FPTAS-general})
of~\cref{th:main-FPTAS}. We have already introduced our smoothness model for
general congestion games in~\cref{sec:smoothed-congestion-games}. Recall that,
under traditional worst-case analysis, computing an $\alpha$-PNE of a congestion
game is \complexclass{PLS}-complete, for any constant
$\alpha$~\parencite{Skopalik2008}. Finally, we emphasize that in general
congestion games we make no monotonicity assumptions, and therefore our results
hold for arbitrary (positive) resource costs. In case one wants to enforce an
increasing assumption (which is common in the literature of congestion games),
the step-function model of the following~\cref{sec:step-function-FPTAS} can be
used instead, in order for the monotonicity to be preserved under smoothness.

Fix an arbitrary congestion game, with $n$ players and $\card{R}=m$ resources.
For simplicity, we will denote the maximum and minimum resource costs by
$$
c_{\max} \coloneqq \max_{r\in R,j\in[n]} c_r(j)
\qquad\text{and}\qquad
c_{\min} \coloneqq \min_{r\in R,j\in[n]} c_r(j).
$$
First observe that, at any outcome $\vecc{s}\in\vecc{S}$, Rosenthal's
potential~\eqref{eq:Rosenthal-potential} can be upper-bounded by
\begin{equation}
\label{eq:potential-upper-bound}
\Phi(\vecc s) = \sum_{r\in R}\sum_{j=1}^{\ell_r(\vecc s)} c_r(j) \leq \sum_{r\in R}\sum_{j=1}^{n} c_r(j) \leq mn \cdot c_{\max}
\end{equation}
and trivially lower-bounded by $\Phi(\vecc s) \geq 0$.
At the same time, the cost of any player $i$ can be lower-bounded by
\begin{equation}
    \label{eq:player-cost-lower-bound}
    C_i(\vecc s) = \sum_{r\in s_i} c_r(\ell_r(\vecc s)) \geq \min_{r\in R} c_r(\ell_r(\vecc s))\geq c_{\min}.
\end{equation}

If $\vecc s \to \vecc s'=(s_i',\vecc{s}_{-i})$ is a move during the execution of
our $(1+\varepsilon)$-BRD, then it must be that $(1+\varepsilon)C_i(\vecc s') <
C_i(\vecc s)$. So, we can lower-bound the improvement of the potential by:
$$
\Phi(\vecc s) - \Phi(\vecc s') 
= C_i(\vecc s) - C_i(\vecc s')
> \frac{\varepsilon}{1+\varepsilon} C_i(\vecc s)
\overset{\eqref{eq:player-cost-lower-bound}}{\geq} \frac{\varepsilon}{1+\varepsilon} c_{\min}.
$$
Therefore, after $T$ steps $\vecc{s}^{0} \to \vecc{s}^{1} \to \dots \to
\vecc{s}^{T}$ of the $(1+\varepsilon)$-BRD it must be that
$$
T \cdot \frac{\varepsilon}{1+\varepsilon} c_{\min} < \Phi(\vecc{s}^{0}) - \Phi(\vecc{s}^{T}) \overset{\eqref{eq:potential-upper-bound}}{\leq} mn c_{\max},
$$
which gives the following upper bound on the number of steps for our dynamics:
\begin{equation}
    \label{eq:bound-BRD-smoothed-ratio-values}
T 
< \left(1+\frac{1}{\varepsilon}\right) nm\frac{c_{\max}}{c_{\min}}.
\end{equation}

On the other hand, recall that our dynamics goes over strategy profiles,
\emph{strictly} decreasing the potential at \emph{every} step; therefore, no two
profiles with the same potential value can be visited via our dynamics.
Additionally, observe that the values $\Phi(\vecc s)$ of Rosenthal's
potential~\eqref{eq:Rosenthal-potential} are fully determined by the
configuration of resource loads $\sset{\ell_r(\vecc s)}_{r\in R}$ under profile
$\vecc s$, and do not directly depend on the actual identities of the players
that use each edge. As a result, we deduce that the total number of iterations
cannot be larger than the number of possible different resource-load profiles.
Since each resource can be used by at most $n$ players, this is at most
$(n+1)^{\card{R}}=(n+1)^m$.
Combining this with \eqref{eq:bound-BRD-smoothed-ratio-values} we can derive
that $(1+\varepsilon)$-BRD terminates after at most
\begin{align}
\label{eq:bound-smoothed-BRD-general-games-helper-1}
T 
&\leq \min\sset{\left(1+\frac{1}{\varepsilon}\right) nm\frac{c_{\max}}{c_{\min}} ,(n+1)^m}
\leq \min\sset{\left(1+\frac{1}{\varepsilon}\right) \frac{nm}{c_{\min}} ,(nm)^m}
\end{align}
iterations. For the last inequality we used the fact that $c_{\max} \leq 1$ and
$n+1\leq nm$, since we can without loss of generality assume that the number of
resources is at least $m\geq 2$; otherwise our congestion game is degenerate,
having only a single strategy profile (and thus the dynamics trivially converge
in constant time).

Since the resource costs $\ssets{c_r(j)}_{r\in R,j\in[n]}$ are independent
$\phi$-smooth random variables, we can now deploy
\cref{lemma:smooth-minimum-order-statistic-bound}, with the choice of parameters
$\mu\gets \card{R}n=mn$, $\beta \gets m$, and $\alpha\gets
\left(1+\frac{1}{\varepsilon}\right)nm$, in order to finally bound the
\emph{expected} number of steps of $(1+\varepsilon)$-BRD
in~\eqref{eq:bound-smoothed-BRD-general-games-helper-1} by
\begin{equation}
    \label{eq:bound-smoothed-BRD-general-games-detailed}
    \phi \cdot \left(1+\frac{1}{\varepsilon}\right) m n \cdot (m+1) \cdot mn\ln(mn) +1 
=
\tilde{O}\left(\frac{1}{\varepsilon}\phi n^2 m^3 \right).
\end{equation}

\subsection{Special Congestion Game Models}
\label{sec:special-games-FPTAS}

\subsubsection{Step-Function Games}
\label{sec:step-function-FPTAS}

In this section we formally describe the model of step-function congestion
games, and prove the corresponding part~(\ref{item:main-FPTAS-step-functions})
of~\cref{th:main-FPTAS}. It is important to mention that the
\complexclass{PLS}-hardness of approximation of~\textcite{Skopalik2008}, that we
have already mentioned for general congestion games, actually uses nondecreasing
costs, so it applies to step-functions as well.

In step-function congestion games, each resource cost function $c_r$ is
represented by a set of $d_r$ integer break points $1=b_{r,1} < b_{r,2} < \dots
<b_{r,d_r}\leq n$ and corresponding value jumps $a_{r,1}, a_{r,2},\dots,
a_{r,d_r}\in (0,1]$. More precisely, the cost of resource $r$ on a load of
$\ell\in[n]$ players is then given by 
$$
c_r(\ell) \coloneqq a_{r,1}+a_{r,2}+\dots + a_{r,\kappa},
\qquad\text{where}\;\; \kappa=\max\sset{j\in[d_r]\fwh{b_{r,j}\leq \ell}}.
$$
Let $d\coloneqq \sum_{r\in R} d_r$ denote the total number of break points across
the entire representation.

For our smoothed analysis, we follow the framework recently proposed
by~\parencite{ggm2022_arxiv} for step-function congestion games, assuming that
the jumps $\ssets{a_{r,j}}$ are independent $\phi$-smooth random variable. We
emphasize, though, that the break points $\ssets{b_{r,j}}$ are not perturbed but
are (adversarially) fixed.

Similarly to
\eqref{eq:player-cost-lower-bound} for general congestion games
(see~\cref{sec:general-games-FPTAS}), if we denote $a_{\min}\coloneqq \min_{r\in
R,j\in[d_r]} a_{r,j}$ and $a_{\max}\coloneqq \max_{r\in R,j\in[d_r]} a_{r,j}$,
the cost of any player $i$, at any strategy profile $\vecc s$,
can be lower-bounded by
\begin{equation}
\label{eq:lower-bounds-potential-player-cost-combined}
\Phi(\vecc s) \geq n a_{\min}
\qquad\text{and}\qquad 
C_i(\vecc s) \geq a_{\min}.
\end{equation}
Also, since all cost functions $c_e$ are now nondecreasing,
from~\eqref{eq:potential-upper-bound} the potential can be upper-bounded as:
\begin{equation}
    \label{eq:upper-bound-potential-step-function}
\Phi(\vecc{s}) 
\leq \sum_{r\in R} n c_r(n)
= n \sum_{r\in R}\sum_{j=1}^{d_r} a_{r,j}
\leq n \sum_{r\in R} d_r a_{\max}
=n d a_{\max}.
\end{equation}

In a totally analogous way
to~\eqref{eq:bound-smoothed-BRD-general-games-helper-1}, following the steps of
the proof of~\cref{sec:general-games-FPTAS}, we can now bound the expected
number of steps of $(1+\varepsilon)$-BRD by
\begin{align*}
    \expect{T} 
    &\leq  \expect{\min\sset{\left(1+\frac{1}{\varepsilon}\right)\frac{nd}{a_{\min}} ,(n+1)^m}}
\end{align*}
Choosing this time parameters $\mu\gets d$, $\alpha\gets \left(1+\frac{1}{\varepsilon}\right) n d$  and $\beta\gets m\frac{\ln (n+1)}{\ln d}$,
we can rewrite this as
\begin{align}
    \expect{T} 
    &\leq \expect{\min\sset{\frac{\alpha}{a_{\min}} ,\mu^{\beta}}} \label{eq:bound-smoothed-BRD-gepolynomial-games-helper-2}.
\end{align}
Noticing that the expectation
in~\eqref{eq:bound-smoothed-BRD-gepolynomial-games-helper-2} is taken with
respect to the independent $\phi$-smooth random variables $\ssets{a_{r,j}}_{r\in
R, j\in[d_r]}$, which are $d=\mu$ many, we can again
deploy~\cref{lemma:smooth-minimum-order-statistic-bound} to finally get a bound
of:
\begin{align*}
    \expect{T} 
&= \tilde{O}\left(\phi\alpha\beta\mu\right)
= \tilde{O}\left(\phi\cdot\left(1+\frac{1}{\varepsilon}\right)nd\cdot m\frac{\ln (n+1)}{\ln d} \cdot d\right)
=\tilde{O}\left(\frac{1}{\varepsilon}\phi nmd^2\right).
\end{align*}

\subsubsection{Polynomial Games}
\label{sec:polynomial-FPTAS}

In this section we introduce the model of polynomial congestion games, and prove
the corresponding part~(\ref{item:main-FPTAS-polynomial})
of~\cref{th:main-FPTAS}. In this model the resource cost functions are
polynomials of a \emph{constant} maximum degree $d$, with nonnegative
coefficients.
It is arguably the most established, and well-studied congestion game model in
algorithmic game theory. Although finding \emph{exact} equilibria in polynomial
congestion games is still a \complexclass{PLS}-complete
problem~\parencite{Ackermann2008,Roughgarden16,ggm2022_arxiv}, no hardness of
approximation results are known. At the same time, the only positive
computational results that we have are for efficiently computing
$d^{O(d)}$-approximate
PNE~\parencite{Caragiannis2011,Feldotto2017,gns2018_journal}. Closing this gap
in our understanding of computability of approximate PNE in polynomial games is
one of the most important remaining open problems in the field.

To continue with the formal definition of our model, each cost function
$c_r$ is represented by a set of coefficients $\ssets{a_{r,j}}_{j\in[0..d]}\in
[0,1]$, where $d\in \N$, so that for all loads $\ell\in[n]$
\begin{equation}
    \label{eq:poly-cost-full}
c_r(\ell)\coloneqq a_{r,0}+ a_{r,1}\ell+\dots + a_{r,d}\ell^{d}.
\end{equation}
We emphasize that the normalization of the coefficients within
$[0,1]$ here is without loss of generality.

To perform smoothed analysis in this congestion game model, the natural choice
is to consider perturbations on the coefficients of the polynomial costs.
However, special care needs to be taken with respect to zero coefficients: any
random noise on them would ``artificially'' introduce monomial terms that did
not exist in the original cost function This, arguably, will distort the
combinatorial aspect of our instance. Therefore, for our smoothness framework we
will assume that only the \emph{nonzero} polynomial coefficients
$\ssets{a_{r,j}}$ are (independent) $\phi$-smooth random variables. 

For that reason, it will be technically convenient to introduce notation
$$
J_r\coloneq\sset{j\in[0..d]\fwh{a_{r,j} > 0}},
\qquad
d_r\coloneqq \card{J_r},
\qquad\text{and}\qquad
\tilde{d} \coloneqq \sum_{r\in R} d_r \leq m(d+1),
$$
for the set of indices of the non-trivial cost coefficients of resource $r$;
then, \eqref{eq:poly-cost-full} can now be written as $ c_r(\ell) = \sum_{j\in
J_r} a_{r,j} \ell^j$. We also let $a_{\min}=\min_{r\in R, j\in J_r} a_{r,j}$ and
$a_{\max}=\max_{r\in R, j\in J_r} a_{r,j}$ for the minimum and maximum nonzero
coefficients across all resources.

It is not hard to verify that we can again derive the same lower bound as
in~\eqref{eq:lower-bounds-potential-player-cost-combined} for the player costs,
and for the upper bound on the potential, due to the monotonicity of the
resource cost function, similarly to
\eqref{eq:upper-bound-potential-step-function} we can now get:
\begin{align*}
\Phi(\vecc s) 
&\leq n\sum_{r\in R} c_r(n)
= n\sum_{r\in R} \sum_{j\in J_r} a_{r,j} n^j
\leq n\sum_{r\in R} d_r a_{\max} n^{d}
= \tilde{d} n^{d+1}  a_{\max}.
\end{align*}

Following along the lines of the derivations for the previous
\cref{sec:general-games-FPTAS,sec:step-function-FPTAS}, we can now bound the
expected number of steps of $(1+\varepsilon)$-BRD by
$$
\expect{T} \leq   \expect{\min\sset{\left(1+\frac{1}{\varepsilon}\right)\frac{\tilde{d} n^{d+1}}{a_{\min}} ,(n+1)^{m}}}.
$$
Choosing parameters $\mu\gets \tilde{d} \leq m(d+1)=O(m)$, $\alpha \gets
\left(1+\frac{1}{\varepsilon}\right)\tilde{d}
n^{d+1}=O\left(\frac{1}{\varepsilon}mn^{d+1}\right)$, and $\beta\gets m\frac{\ln
(n+1)}{\ln \tilde{d}}=\tilde{O}(m)$, similarly to the proof
in~\cref{sec:step-function-FPTAS} we can again derive the bound
in~\eqref{eq:bound-smoothed-BRD-gepolynomial-games-helper-2}. Thus,
applying~\cref{lemma:smooth-minimum-order-statistic-bound} for the
$\tilde{d}=\mu$ many independent $\phi$-smooth random variables
$\ssets{a_{r,j}}_{r\in R, j\in J_r}$, we can now bound the expected number of
steps of our dynamics by:
\begin{align*}
    \expect{T} 
&=  \tilde{O}\left(\phi\alpha\beta\mu\right)
=\tilde{O}\left(\phi \cdot \frac{1}{\varepsilon}mn^{d+1} \cdot m \cdot m\right)
=\tilde{O}\left(\frac{1}{\varepsilon} \phi n^{d+1} m^3\right).
\end{align*}

\subsubsection{Cost-Sharing Games}
\label{sec:cost-sharing-FPTAS}

This section deals with the proof of part~(\ref{item:main-FPTAS-cost-sharing})
of~\cref{th:main-FPTAS}. 
\emph{Fair cost-sharing} games are congestion games where the cost of resources
are given by $c_r(\ell)=\frac{a_r}{\ell}$, where $a_r>0$. Notice that these are
\emph{decreasing} functions; they can be interpreted as a fixed edge cost $a_r$
being equally split among the players that use it. It is known that the problem
of finding an exact PNE in fair cost-sharing games is
\complexclass{PLS}-complete, even in network games~\parencite{Syrgkanis2010},
and that better-response dynamics take exponentially long to
converge~\parencite{ADKTWR04}. To the best of our knowledge, no positive results
exist regarding the efficient computation of \emph{approximate} PNE, apart from
the very special case of metric network facility location games, with uniform
costs~\parencite{Hansen2009}.

Unlike the previous congestion game models studied in this paper, no smoothness
framework has been proposed before for cost-sharing games. Therefore, we propose
here to consider $a_r$ as independent $\phi$-smooth random variables; arguably,
this seems as the most natural approach.

Like before, we denote $a_{\min}\coloneqq \min_{r\in R} a_r$ and
$a_{\max}\coloneqq \max_{r\in R} a_r$. Given that resource costs are now
decreasing, we can lower-bound the player costs at any profile $\vecc s$ by
$$C_i(\vecc s)\geq \min_{r\in R} c_r(\ell_r(\vecc s))\geq \min_{r\in
R}c_r(n)=\min_{r\in R}\frac{a_e}{n}=\frac{1}{n}\cdot a_{\min}.$$
Furthermore, we can upper-bound the potential values by 
\begin{align*}
    \Phi(\vecc s) &=\sum_{r\in R}\sum_{j=1}^{\ell_r(\vecc s)} c_r(j)\leq m \max_{r\in R} \sum_{j=1}^n c_r(j) = m \max_{r\in R}\sum_{j=1}^n\frac{a_r}{j}= m H_n \cdot a_{\max},
\end{align*}
where $H_n\coloneqq \sum_{j=1}^n\frac{1}{j}$ is the harmonic numbers function.

Using the above inequalities, we get the following bound on the expected number
of steps of $(1+\varepsilon)$-BRD, analogously
to~\eqref{eq:bound-smoothed-BRD-general-games-helper-1}:
$$
\expect{T}
\leq   \expect{\min\sset{\left(1+\frac{1}{\varepsilon}\right) \frac{nmH_n}{a_{\min}} \ln\left(\frac{m}{a_{\min}}\right),(n+1)^{m}}}.
$$
Similarly to our derivation in~\cref{sec:step-function-FPTAS}, we can
deploy~\cref{lemma:smooth-minimum-order-statistic-bound}, this time with
parameters $\mu \gets m$, $\alpha \gets
\left(1+\frac{1}{\varepsilon}\right)nmH_n=O\left(\frac{1}{\varepsilon}mn\log{n}\right)=\tilde{O}\left(\frac{1}{\varepsilon}mn\right)$
and $\beta \gets m\frac{\ln(n+1)}{\ln m}=\tilde{O}(m)$, to get the following
bound
\begin{align*}
\expect{T}
&=\tilde{O}\left(\phi\alpha\beta\mu\right)
= \tilde{O}\left(\phi\cdot \frac{1}{\varepsilon}mn \cdot m \cdot m\right)
= \tilde{O}\left(\frac{1}{\varepsilon}\phi n m^3\right).
\end{align*}

\bigskip
\bigskip

\paragraph*{Acknowledgments} I am truly grateful to the anonymous reviewer of
EC'24 who suggested the current, simplified approach over a previous version of
the analysis in~\cref{sec:general-games-FPTAS}, and which resulted in an
improved bound in~\eqref{eq:bound-BRD-smoothed-ratio-values}, saving a
logarithmic factor. The same reviewer also suggested a modification in the proof
of~\cref{lemma:smooth-minimum-order-statistic-bound}, that resulted in an
improvement in the bound in~\eqref{eq:smooth-minimum-order-statistic-bound} as
well. This allowed us to improve some  powers of the polynomial factors in all
our smoothed running time bounds. 

\newpage
\printbibliography

\end{document}